# Theoretical simulation and design of AlSb thin films solar cells*

Huijin Song[a], Zilong Wang[a], Jingwen Wang[a], Qiang Yan[a,b*], Xia Kai[a], Xiangfeng Deng[a], Minqiang Li[c]

[a] College of Mechanical Engineering, Chengdu University, Chengdu 610106,China;
[b] Sichuan Yuqiang Science and Technology Co., Ltd. ,Mianzhu 618200,China;
[c] School of Electronic Engineering,Chengdu Technological University, Chengdu, 610031, China

**Abstract**：The effects of thickness, doping concentration and recombination of AlSb films on the performance of CdS/AlSb cells are simulated by one dimensional simulation program called analysis of microelectronic and photonic structures(AMPS1D) soft ware to understand the influence of material characteristic (such as carrier concentration and thickness) on the solar cells. The methods to improve the performance of CdS/AlSb cells by optimizing the properties of AlSb have been found. The results show that the thicker AlSb film can improve the long wave response for the higher short-circuit current density ($J_{sc}$) of CdS/AlSb solar cells and the higher carrier concentration of the film can improve open-circuit voltage ($V_{oc}$) and fill factor (*FF)*, and its optical thickness for CdS/AlSb solar cells is in the range of 500nm~2000nm. The conversion efficiency can be improved from 10.6% to15.3% for introducing AlSb:Te, AlSb:Cu and ZnTe:Cu thin films to CdS/ AlSb structure. Furthermore, the thicker AlSb:Te film can improve the short wave response for the higher $J_{sc}$ of the cells, and its optical thickness CdS/AlSb:Te/AlSb/ZnTe:Cu solar cells is in the range of 100nm~200nm. And the lower doping concentration can promote $V_{oc}$ and *FF* to improve the characteristic of the cells.

# 1 Introduction

Low cost and high efficiency thin film solar cells have been paid more and more attention with the development of solar cells[1].Aluminium antimonide(AlSb) with the sphalerite structure has optical indirect band gap width of 1.62 eV [2-4], which is agreement with the visible light spectrum. The conversion efficiency of AlSb solar cells in theoretic can be as high as 27% [5,6]. AlSb is promising for applications in high temperature applications, p-n junction diodes and high energy photon detectors [7].

AlSb film photoelectric materials are environment-friendly and exhibited potential application for the abundant Al and Sb in the earth and non-toxic in manufacturing and use[8,9]. But there are a few reports on current research for the difficult preparation and easy deliquesce of AlSb poly crystalline film[8,10]. Johnson J E obtained AlSb thin films by simultaneously depositing Al and Sb elements on a 550 ℃ substrate[11]. Lal K and Singh T prepared AlSb thin films by single crystal evaporation [12] and hot wall epitaxial method [13]. J.C.He prepared AlSb thin films by magnetron sputtering and studied the properties[14]. Singh M et al. firstly coated an Al film with a layer of Sb film by thermal evaporation, and obtained an annealing method. P-type AlSb film [15]. In the early stage, AlSb polycrystalline thin films were prepared by annealing by co-evaporation method. The structure and properties of the films were studied. TCO/CdS/AlSb solar cells with open circuit voltage of about 200mV were obtained [16]. However, regarding the preparation of AlSb solar cells, relevant international reports are rare.

In order to provide theoretical guidance for the preparation of AlSb solar cells, the energy band structure was analyzed and the built-in electric field distribution of AlSb solar cells was discussed by using AMPS-1D software from the discussion of the physical model and mathematical solution of solar cells.

2 Basic equations and solutions for solar cells

The basic equation of a solar cell consists of a Poisson equation that satisfies the charge space distribution and a continuous equation that the carrier transport needs to satisfy. which is

$$\frac{d}{dx}\left[\varepsilon(x)\frac{d\psi}{dx}\right] = q\cdot\left[p(x)-n(x)+N_D^+(x)-N_A^-(x)+p_t(x)-N_t(x)\right] \quad (1)$$

$$\frac{dJ_n}{dx} = q(-G_{op}(x)+R(x)) \quad (2)$$

$$\frac{dJ_p}{dx} = q(G_{op}(x)+R(x)) \quad (3)$$

In the formula, $\psi(x)$ is the partial vacuum level, $n(x)$, $p(x)$ are the trapped electron concentration and hole concentration, respectively. $N_{D+}(x)$ and $N_{A-}(x)$ are ionized donor and class acceptor doping concentrations, respectively. $\varepsilon(x)$ and q are dielectric constant and electron charge, respectively. $J_n(x)$ and $J_p(x)$ are current densities of free electrons and free holes, respectively. $G_{op}(x)$ and $R(x)$ are the photogenerated carrier generation rate and recombination rate, respectively.

Set the bias voltage to V, at the boundary of the device x=0 and x=L,

$$\psi(0) = \psi_0 - V \quad (4a)$$

$$\psi(L) = 0 \quad (4b)$$

$$J_p(0) = qS_{p0}(p(0)-p_0(0)) \quad (4c)$$

$$J_p(L) = qS_{pL}(p(L)-p_0(L)) \quad (4d)$$

$$J_n(L) = qS_{n0}(n(0)-n_0(0)) \quad (4e)$$

$$J_n(L) = qS_{nL}(n(L)-n_0(L)) \quad (4f)$$

The subscripts 0 and L represent the left and right boundaries of any device. Take the vacuum level $\psi(L)$ of x=L as the energy level reference point. Under thermal equilibrium, $\psi(0)=0$, $\psi(L)=0$. When there is bias or light, $\psi(0)=\psi_0-V$.

$S_{n0}$, $S_{nL}$, $S_{p0}$, and $S_{pL}$ are the effective surface recombination velocities of electrons and holes at x=0 and x=L, respectively. Through the selection of the S value, different transport processes can be characterized. For example, $S_{p0}$ is considered as the thermal activation speed of the cavity, and the hole is determined by the thermal excitation process through the x=L process. S represents the surface recombination velocity, and the magnitude of which reflects the degree of surface passivation. The S value and the barrier height of the electrode before and after x=0 and x=L are selected, that is, the contact characteristics of the front and rear electrodes can be characterized. For example, to obtain an ideal ohmic contact electrode at x = 0, then $S_{n0}$ is chosen to be large enough to ensure that n(0) = n0(0) holds true under any bias conditions.

Thus, if measurable variable expressions of n, p, $N_{d+}$, $N_{v-}$, $n_t$, $p_t$, $G_{op}(x)$, and R(x) are given, the values of these variables are measured, and electronic static electricity can be calculated by numerical calculation. Potential energy ψ, to calculate the corresponding carrier concentration distribution, space charge density concentration distribution, electric field intensity distribution and energy band diagram.

The equations in the basic equations of the solar cell are nonlinear and related to each other and can only be solved numerically.

When solving, first divide the device into several regions, set N, N adjacent regions and end faces together, for a total of $N^{+1}$ boundaries.

For each region divided, the continuous equations of electron current and hole current are expressed as:

$$f_{ei}(x) = \frac{2}{q(h+H)}\left[j_{n,i+\frac{1}{2}}(x) - j_{n,i-\frac{1}{2}}(x)\right] + G_{op\,i}(x) - R_i(x) \quad (5a)$$

$$f_{hi}(x) = \frac{2}{q(h+H)}\left[ j_{p,i+\frac{1}{2}}(x) - j_{p,i-\frac{1}{2}}(x) \right] + G_{op\,i}(x) - R_i(x) \tag{5b}$$

$G_{op}$ is the photon flux density of the device, the photon energy $h_{vi}=E_{gopt}$, and the photon yield at any position as the photon moves from x=0 to x=L:

$$G_{op}(x) = -\frac{d}{dx}\sum_i \Phi_i^{FOR}(x) + \frac{d}{dx}\sum_i \Phi_i^{REV}(x) \tag{5c}$$

The superscript FOR represents the photon flow from x=0 to x=L in the direction of travel, and REV represents the photon flow from x=L to x=0, which is reflected back by the inner surface or end face of the device.

R(x) is the sum of carriers' direct transition complex ($R_D$) between bands and the indirect transition complex ($R_I$) with band gap states.

This will result in 3N-3 linearized equations and 6 boundary conditions of equations (4a)-(4f) constituting a system of 3N+3 equations. The equations are linearized by the Newton-Rapson method to obtain an iterative format.

$$[A]\bullet[\delta] = [B] \tag{5d}$$

among them

$$[A] = \begin{bmatrix} \Uparrow \\ \psi \\ E_{fpi} \\ E_{fni} \\ \Downarrow \end{bmatrix} \quad [B] = \begin{bmatrix} \Uparrow \\ f \\ E_{ei} \\ E_{hi} \\ \Downarrow \end{bmatrix} \quad [\delta] = \begin{bmatrix} \Uparrow \\ \delta\psi \\ \delta E_{fpi} \\ \delta E_{fni} \\ \Downarrow \end{bmatrix},$$

The matrix represents the difference between physical quantities such as $\psi$, $E_{fpi}$, and $E_{fni}$ before and after the iteration. In the calculation, first calculate the $\psi$, $E_{fpi}$, $E_{fni}$ in the thermal equilibrium state, and use this as the initial amount to calculate the parameters such as $\psi$, $E_{fpi}$, $E_{fni}$, electron

current,and hole current in the case of adding voltage or light,thereby obtaining the current-voltage characteristics of the device.From the $j_n(x) = q\mu_n n(dE_{fn}/dx)$ and $j_p(x) = q\mu_p n(dE_{fp}/dx)$ get the electron current and the hole current,the electric field distribution in the device is obtained from dE/dx.The recombination rate R,the carrier concentration n,p,and the charged defect state density $n_t$,$p_t$ can be calculated from the corresponding expressions.

## 3 AlSb polycrystalline thin film solar cell device

In this paper, AMPS-1D simulation software was used to simulate and analyze the characteristics of AlSb solar cells. And the internal electric field, band structure, I-V characteristics, carrier recombination of the cells were obtained. The effects of film properties on device characteristics were discussed and compared. The material parameters used in the simulation were determined based on the results reported in the literature and the actual measurements.The specific parameters were shown in Tab.1.

Tab.1 Physics parameters in the simulation

| | Thickness (μm) | Dielectric constant | Electronic affinity (eV) | Energy gap (eV) | Carrier concentration (cm$^{-3}$) | Electron mobility (cm$^2$/V.Sec) | Hole mobility (cm$^2$/V.Sec) | Life (Sec) |
|---|---|---|---|---|---|---|---|---|
| n-CdS | 0.15 | 9 | 4.5 | 2.42 | $1\times10^{17}$ | 340 | 50 | $10^{-9}$ |
| n-AlSb:Te | variable | 12.04[17] | 3.6 | 1.62 | variable | 200[18] | 300[19] | $10^{-8}$ |
| AlSb | variable | 12.04[18] | 3.6[17] | 1.62 | variable | 200[18] | 300[19] | $10^{-8}$ |
| ZnTe:Cu | 0.1 | 10.1[19] | 3.53[19] | 2.26 | $1\times10^{20}$ | 50[19] | 50[19] | $10^{-9}$ |

3.1 Effect of the thickness of AlSb film on the performance of AlSb solar cells

Fig.1 is an energy band diagram of the CdS/AlSb basic structure.The thickness of the AlSb film was 0.5 μm, 1 μm, 2 μm, 4 μm and 6 μm, respectively, and the carrier concentration of the AlSb layer was $1\times10^{17}$ cm$^{-3}$. The remaining parameters are given in Tab.1. At the set carrier concentration, the contact barrier of the front electrode and the back electrode is 0.02 eV and 0.003 eV, which is close to the ideal ohmic contact, so the influence of the carrier concentration on the performance of CdS/AlSb solar cells can be compared.

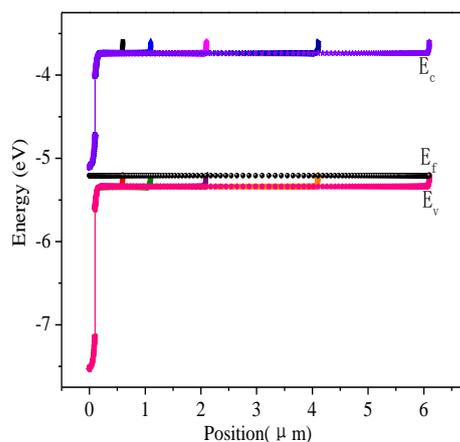

Fig.1 the band structure of CdS/AlSb solar cells with different thickness of AlSb films

（Supposing front and back electrode barrier were 0.02eV and 0.03eV, respectively.）

Tab.2 the performance parameters of CdS/AlSb solar cells with different thickness of AlSb films

| Thickness(μm) | $V_{oc}$ (V) | $J_{sc}$ (mA/cm$^2$) | FF | η (%) |
|---|---|---|---|---|
| 0.5 | 1.092 | 17.592 | 0.554 | 10.639 |
| 1 | 1.088 | 17.245 | 0.545 | 10.217 |
| 2 | 1.087 | 17.130 | 0.543 | 10.120 |
| 4 | 1.087 | 17.116 | 0.543 | 10.112 |
| 6 | 1.087 | 17.116 | 0.543 | 10.112 |

Fig.2 shows the simulation results of the light *J-V* characteristics of CdS/AlSb solar cells. The performance parameters of the battery were calculated as shown in Tab.2. It can be seen that the reduction of the thickness of the AlSb film is beneficial to the improvement of the performance of the device in various aspects, especially the current density. From the spectral response curve (Fig.3), it can be seen that the increase of the short-circuit current density mainly comes from longer wave above 500 nm.Overall, the conversion efficiency of the battery decreases slightly with the thickness of AlSb increasing.The physical mechanism is: for the solar cells where $V_{oc}$ has basically become saturated, by the formula $\eta = \dfrac{V_{oc} I_{sc} FF}{A_t P_{in}} = \dfrac{V_{oc} J_{sc} FF}{P_{in}}$ ($A_t$ is the cell area, $P_{in}$ is the solar intensity per unit area), and the change in $\eta$ is mainly affected by $J_{sc}$.As the thickness of the AlSb film gradually increases in the same order of magnitude, the series resistance of the cell slightly increases, so the $J_{sc}$ gradually decreases, resulting in a slight decrease in the conversion efficiency $\eta$.

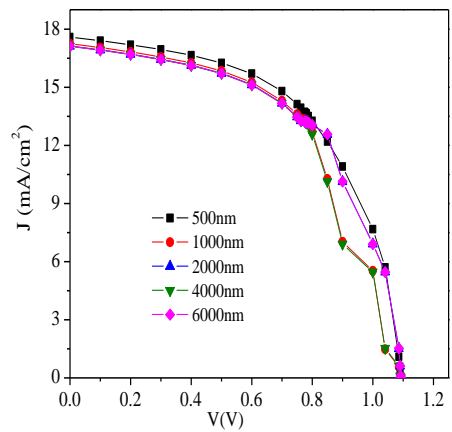

Fig.2 J-V curves of CdS/AlSb solar cells with different thickness of AlSb films

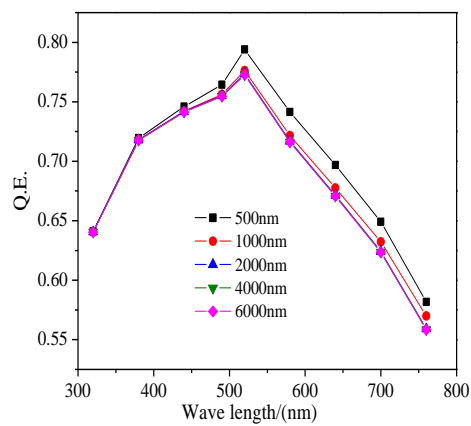

Fig.3 Spectral response curve of CdS/AlSb solar cells with different thickness of AlSb films

Therefore, according to the above results, considering the electric field distribution and the series resistance of the cells, the thickness of the AlSb film is between 0.5 μm and 2 μm while ensuring that the device is guaranteed not to be short-circuited for the desired device performance,

3.2 Effect of carrier concentration of AlSb film on the performance of AlSb solar cells

Fig.4 and Fig. 5 show the simulation results of the carrier concentration of the AlSb film and the performance of the AlSb solar cell, respectively. The cell structure is CdS/AlSb. Tab.3 shows the corresponding battery device characteristic parameters.In the simulation, in order to simplify the calculation, it is considered that the back electrode contact barrier is 0.3 eV at different carrier concentrations.When the actual battery is constant in the back electrode metal, and the work function of the metal is lower than the that of AlSb, the back contact barrier becomes larger as the carrier concentration increases. But when the carrier concentration is sufficiently high, AlSb and The depletion region of the Schottky junction formed by the back electrode metal will become sufficiently narrow that the tunneling current will be significantly higher than the thermal emission current and become the main component of the current.The simulation calculation procedure does not consider tunnel transport, so the simplified back contact barrier remains unchanged at 0.3 eV.The results show that the short-circuit current density decreases with the increase of carrier concentration, and the open circuit voltage increases with the increase of carrier concentration.The physical mechanism is: the increase of carrier concentration makes the depletion region of the AlSb near the back electrode narrow and the probability of transporting holes to the back electrode increase. So that the series resistance of the battery decreases and the filling factor of the battery increases with the concentration of the carrier increasing although the back contact barrier is unchanged。

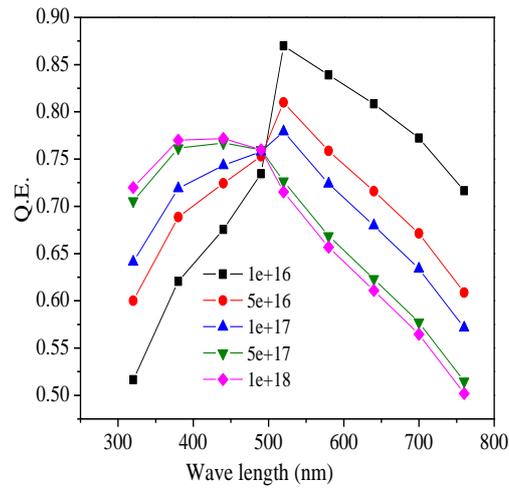

Fig.4 QE curves of CdS/AlSb solar cells with different hole concentration of AlSb films

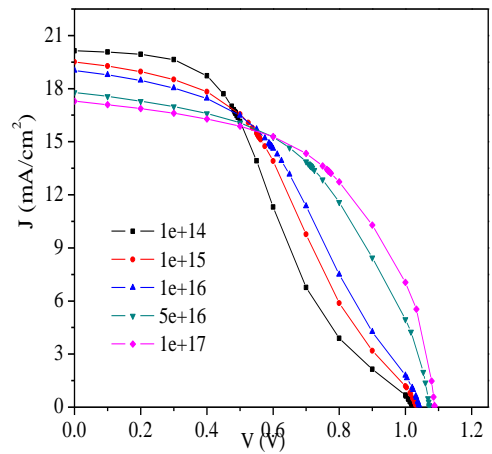

Fig.5 Spectral response curves of CdS/AlSb solar cells with different hole concentration of AlSb films

Tab.3 Performance parameters of CdS/AlSb solar cells with different hole concentration of AlSb films

| Hole concentration (cm$^{-3}$) | $V_{oc}$ (V) | $J_{sc}$ (mA/cm$^2$) | FF | $\eta$ (%) |
| --- | --- | --- | --- | --- |
| $1\times10^{14}$ | 1.021 | 20.13 | 0.393 | 8.09 |
| $1\times10^{15}$ | 1.032 | 19.51 | 0.423 | 8.51 |
| $1\times10^{16}$ | 1.042 | 19.03 | 0.442 | 8.77 |
| $5\times10^{16}$ | 1.073 | 17.79 | 0.510 | 9.73 |
| $1\times10^{17}$ | 1.088 | 17.29 | 0.544 | 10.24 |

3.3 AlSb: Effect of thickness of Te film on properties of AlSb solar cells

Based on the improvement of lattice mismatch, energy band and AlSb film properties of the devices, the thickness of AlSb:Te film was calculated for CdS/AlSb:Te/AlSb/AlSb:Cu/ZnTe:Cu solar cell performance.The film parameters are given in Tab.1.

Fig.6, 7 and 8 are the built-in electric field diagram, energy band diagram and spectral response diagram of CdS/AlSb:Te/AlSb/AlSb:Cu/ZnTe:Cu solar cells with different thickness of AlSb:Te films. Comparing Fig.1, a back junction is added to the front p-n junction to form a built-in electric field as the pn junction in the same direction after introducing the ZnTe:Cu back contact layer to the AlSb solar cells, which is beneficial to improve the collection of hole carriers.It can be seen that reducing the film thickness is beneficial to the performance of the device in various aspects (see Tab.4), especially for the short-circuit current density.It is seen from the spectral response curve (Fig. 9) that the increasing of the short-circuit current density is mainly the contribution of the short-wave below 550 nm.Therefore, reducing the thickness of the AlSb:Te film is a way to improve the efficiency, but the reducing the AlSb:Te polycrystalline film needs to be performed without ensuring that the device is not short-circuited.

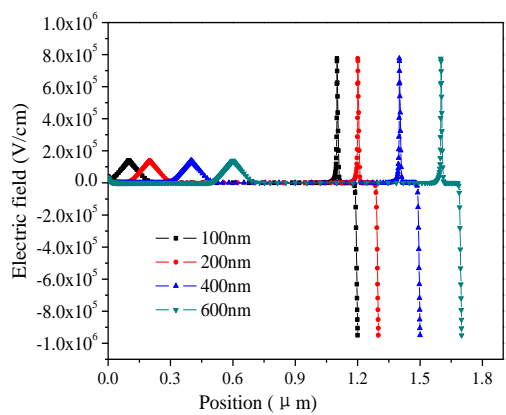

Fig.6 the interior electric field of CdS/AlSb:Te/AlSb/AlSb:Cu/ZnTe:Cu solar cells with different thickness of AlSb:Te films

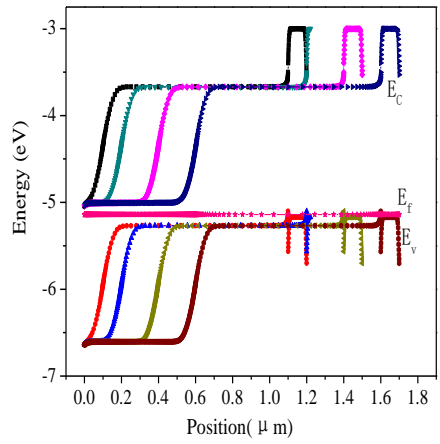

Fig7.the band structure of CdS/AlSb:Te/AlSb
/AlSb:Cu/ZnTe:Cu solar cells with different thickness of AlSb:Te films
（Supposing front and back electrode barrier were 0.02eV and 0.03eV, respectively.）

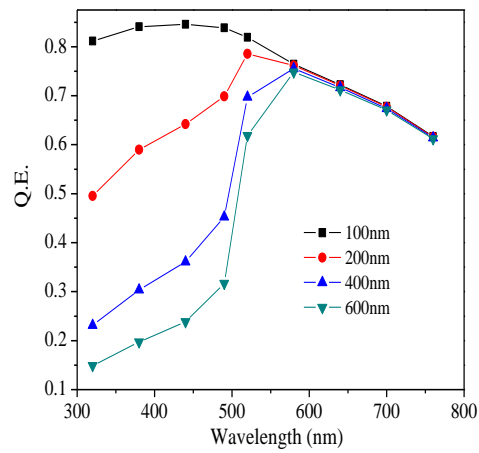

Fig.8 Spectral response curves of CdS/AlSb:Te /AlSb/AlSb:Cu/ZnTe:Cu solar cells with different thickness of AlSb:Te films

Tab.4 the performance parameters of CdS/AlSb:Te/AlSb/AlSb:Cu/ZnTe:Cu solar cells different thickness of AlSb:Te films

| Thickness (nm) | $V_{oc}$ (V) | $J_{sc}$ (mA/cm$^2$) | FF | $\eta$ (%) |
| --- | --- | --- | --- | --- |
| 100 | 1.032 | 18.785 | 0.789 | 15.300 |
| 200 | 1.044 | 17.314 | 0.785 | 14.189 |
| 400 | 1.053 | 15.168 | 0.797 | 12.719 |
| 600 | 1.057 | 14.084 | 0.801 | 11.929 |

### 3.4 Effect of Carrier Concentration of Te Thin Film on Properties of AlSb Solar Cells

For our experiments, we found that the carrier concentration of AlSb:Te film is $10^{16} \sim 10^{17} cm^{-3}$ [20], and we carried out device performance simulation. Tab.5 gives the simulated performance parameters. It can be seen that the conversion efficiency and fill factor of the battery decrease as the carrier concentration increases.

Tab.5 the performance parameters ofCdS/AlSb:Te/AlSb/AlSb:Cu/ZnTe:Cu solar cells with hole concentration of AlSb:Te films

| Hole concentration (cm$^{-3}$) | $V_{oc}$(V) | $J_{sc}$ (mA/cm$^2$) | FF | $\eta$ (%) |
|---|---|---|---|---|
| $1 \times 10^{16}$ | 1.018 | 18.64 | 0.811 | 15.39 |
| $5 \times 10^{16}$ | 1.023 | 20.05 | 0.738 | 15.13 |
| $1 \times 10^{17}$ | 1.032 | 18.78 | 0.789 | 15.30 |
| $5 \times 10^{17}$ | 1.059 | 17.63 | 0.798 | 14.89 |

## 4 Summary

In this paper，the effect of AlSl and AlSb:Sb films on solar cells were systematically simulated. It was found that the thinner AlSb films improved the long-wave response of CdS/AlSb Solar Cells.The effect of Te film on battery performance gives ways to improve battery performance. The main conclusions as following:

(1) The thinning of the AlSb film improves the long-wave response of the CdS/AlSb solar cell, thereby increasing the short-circuit current density and improving the conversion efficiency of the solar cells. Therefore, based on ensuring that the AlSb film does not have blisters and completely covers the CdS film, the thickness of the AlSb film is reduced and maintained between 500 nm and 2000 nm to improve the efficiency of the solar cells. An increasing of the carrier concentration of the AlSb film will improve the fill factor and open circuit voltage to improve AlSb solar cells' performance.

(2) Reducing the thickness of AlSb:Te films improves the short-wave response and the short-circuit current density, to improve the device performance. Therefore, the thickness of the AlSb:Te film in the CdS/AlSb:Te/AlSb/AlSb:Cu/ZnTe:Cu solar cell is maintained between 100 nm and 200 nm. Decreasing the carrier concentration of AlSb:Te film will improve the fill factor and open circuit voltageto improve device performance.

## 参考文献

## Acknowledgment

This project was supported financially by the Open Research Subject of Powder Metallurgy Engineering Technology Research Center of Sichuan Province (Grant Nos. SC-FMYJ2019-06) and the Training Program for Innovation of Chengdu University, China (Grant No. 201911079003).